# Gamma radiation exposure of MCT diode arrays

**F F Sizov[1], I.O. Lysiuk, J V Gumenjuk-Sichevska, S G Bunchuk, V V Zabudsky**

Institute of Semiconductor Physics, 03028, Nauki Av. 41, Kiev, Ukraine,
e-mail: [1]*sizov@isp.kiev.ua*

**Abstract.** Investigations of electrical properties of long-wavelength infrared (LWIR) mercury cadmium telluride (MCT) arrays exposed to γ-radiation have been performed. Resistance-area product characteristics of LWIR $n^+$-$p$-photodiodes have been investigated using microprobe technique at T ≈ 78 K before and after an exposure to various doses of γ-radiation ($Co^{60}$ Gammas). The current transport mechanisms for those structures are described within the framework of the balance equation model taking into account the occupation of the trap states in the band gap.

**Keywords**: MCT arrays, carrier transport, radiation hardness.

## 1. Introduction

IR imagers at the moment are being increasingly used in different kinds of applications for passive night-vision observations, tracking, etc. In contemporary systems the key components are multielement linear scanning or matrix arrays. Among the various materials used for the fabrication of IR arrays, mercury cadmium telluride (MCT) is currently the most widely used material both for (3 – 5) μm and (8 – 14) μm atmospheric transparency windows.

In spite of the complications involved in the physics of MCT devices (intensive Shockley-Read-Hall recombination and trap-assisted tunnelling due to a large number of different types of defects in the gap) HgCdTe moderately cooled arrays remain the most sensitive small pitch FPAs, e.g., compared to quantum well infrared photodetectors (QWIP) (both AlGaAs QWIPs and GaInSb strain layer superlattices), doped silicon detectors, uncooled microbolometer arrays, InSb photodiodes, Schottky-barrier detectors, etc.) in long-wavelength infrared (LWIR) and middle-wavelength infrared (MWIR) regions operating at high frame rates and low integration time. None of these competitor arrays can rival MCT FPAs in terms of fundamental properties. The arrays of other materials mentioned seem in some aspects more manufacturable, but cannot achieve higher performance at comparable temperatures of operation and even at lower temperatures. Now HgCdTe solid solution is nearly ideal material for photovoltaic (PV) infrared detectors [1] both for MWIR (3 to 5 μm) and LWIR (8 to 12 μm) infrared regions. It is important to ascertain the functionality of arrays manufactured in MCT MBE grown epitaxial layers under exposure to the radiation environment and try to disclose the physics of the process. Earlier similar investigations applied to some kind of MCT devices were carried out in various radiation environments (neutrons, electrons, protons) [2].

There exist several parameters for FPAs to be characterized: sensitivity, detectivity, noise equivalent temperature difference (NETD) and some others. But initially most of them are based on R·A-products obtained from current-voltage (C-V) characteristics of IR-photodiodes array, their photoresponse and proper signal handling by charged-coupled device (CCD) or complementary metal-oxide-semiconductor (CMOS) -readouts.

The γ-radiation influence on electrical properties of MBE HgCdTe films was studied in [3,4] and found to be within the range of experimental accuracy. The purpose of this investigation was to get data about γ-radiation influence on the properties of $p$-$n$-junctions in the arrays prepared in similar HgCdTe films, grown by MBE method at the same conditions with cap $Hg_{1-x}Cd_xTe$ layers with grad chemical composition up to x ≈ 0.35, which is less that in CdTe layers earlier used as stable protective layers for MCT photodiodes [2].

In the present work we use current-voltage characteristics to determine the changes of carrier transport in arrays of MCT $n^+$-$p$-junctions for the LWIR ($\lambda$ ≈ 8-12 μm) spectral region after an exposure to γ-radiation.

## 2. Radiation hardness of MCT arrays

For IR arrays applications, the issue of radiation stability is important from the point of view of the upper possible dose limits of their operation capability. It puts certain requirements to the fabrication methods and design of the hybrid arrays as well.

In the present work, dark current carrier transport and radiation hardness was studied in MCT photodiodes prepared from the layers grown by molecular beam epitaxy (MBE) [5,6]. All elements of the diode array were tested by an electrical microprobe with the edge diameter of 10 μm and nitrogen backpressure at 78 K.

MBE HgCdTe epitaxial layers were grown on 2-inch-diameter (013) GaAs substrates with an intermediate CdZnTe buffer layer [7]. The growth temperature was within $T = 240 \div 300^0C$ for CdZnTe buffer layers and within $T \approx 180 \div 190^0C$ for the HgCdTe layers. The inactive layer of HgCdTe had the gradient $x$ which changed from 0.35 at the surface to the constant value of 0.2 at the depths over 1 μm. During the growth process the composition of the layer was controlled by a built-in ellipsometer. The composition non-uniformity over the area of about 1 cm$^2$ did not exceed $\Delta x = \pm 0.001$. The as-grown layers were of $n$-type conductivity (electron concentration $n_{78} \approx 2 \cdot 10^{15}$ cm$^{-3}$, electron mobility $\mu_{78} \approx 10^5$ cm$^2$/V×sec). Annealing was necessary to convert the $n$-conductivity type of the layers into the $p$-type. After the annealing procedure the layers had hole concentrations about $p_{78} = 6.3 \cdot 10^{15}$ cm$^{-3}$ with carrier mobility in the range of $\mu_{78} \approx 550$ cm$^2$/V·s. The subsequent boron implantation was used to produce $n^+$-$p$-diodes to be used with $n$-channel CCD or CCD + CMOS readouts [8]. The dislocation densities in the material for those IR photodetectors were about 10$^6$ cm$^{-2}$ (as determined by the etch pit density method).

Co$^{60}$ was used as a source of gamma radiation. C-V characteristics of photodiode arrays have been measured before and after γ-radiation exposures of 1·10$^4$ R, 1.1·10$^5$ R, 1.1·10$^6$ R, and 1.46·10$^7$ R. It was found that two groups of photodiodes in arrays can be distinguished. Most of the pixel elements in the arrays (near 70%) did not change their operational characteristics such as C-V-characteristics and differential resistance, even when exposed to γ-radiation with the dose up to 1.46·10$^7$ R as shown for a typical diode in Figure 1.

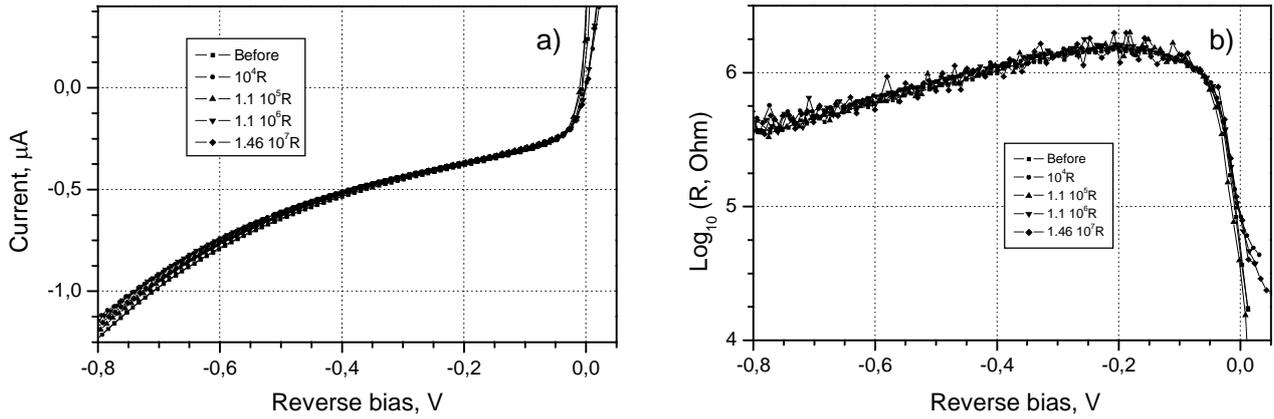

Figure 1. Typical experimental dependences of dark currents and differential resistances on bias for the majority (≈70%) of $Hg_{1-x}Cd_xTe$ $n^+$-$p$-photodiodes which are not changing their properties when exposed to γ-radiation, $T$=78 K.

In the other group of the diodes, slightly decreased incremental resistances and larger currents after influence of γ-radiation were observed as shown in Figure 2. In this group of diodes, changes of currents and differential resistances are not linearly dependent on the dose. After the exposure of $10^4$ R the C-V characteristics are sharply deteriorating. However, subsequent treatments with an increased dose lead to a gradual improvement of the dark current: at the dose of $1.46 \cdot 10^7$ R the dark current is approximately twice as high as for unexposed biased diodes. It should be pointed out that the initial characteristics for both groups of the diodes were practically identical.

At the highest radiation dose the level of noise in contacts is increased for both groups of the diodes, as seen from differential resistance characteristics in Figures 1b, 2b. This is caused by the fact that the indium contacts oxidize under the γ-radiation exposure.

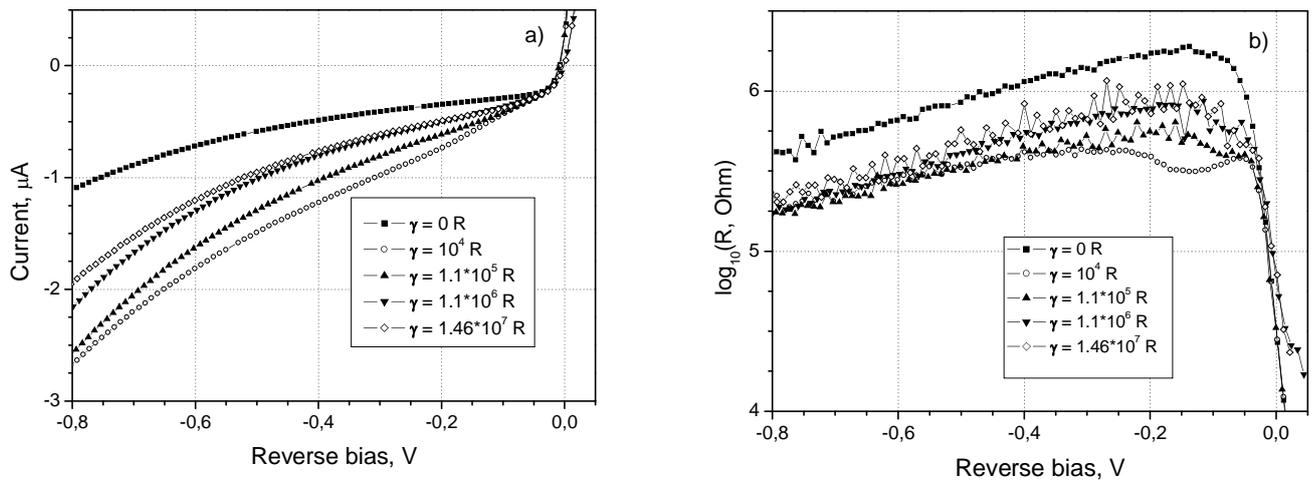

Figure 2. Typical experimental dependence of dark current and differential resistances on bias for $Hg_{1-x}Cd_xTe$ $n^+$-$p$-photodiodes that exhibit a significant response to γ-radiation treatment, $T$=78 K.

As it was shown in [3,4], stable donor-type radiation defects are formed in MCT layers under the influence of high-energy β- and γ-radiation. Therefore, the radiation damage of the structure can be estimated from the dependence of the active defects concentration on the radiation dose. In the present paper, we fit the experimental C-V-curves with the help of the balance equations method (successfully used earlier [9,7,14] for the calculation and optimization of photodiode parameters) in order to extract information on the changes of electrically active defects concentration caused by radiation.

### 3. Balance equation method for $n^+$-$p$-junctions and the fitting procedure

For investigated diodes with a cut-off wavelength $\lambda_{co}$ = 10.3 ± 0.3 µm and the values of dynamic resistance within $R_0A$ ≈ 10-20 Ohm·cm$^2$, the main contribution to dark current via $n^+$-$p$-junctions with concentrations of $p$-type layers of $p_{77}$ ≈ (5÷10)×10$^{15}$ cm$^{-3}$ comes from diffusion, generation-recombination trap assisted tunnelling processes, and from band-to-band tunnelling [9].

The band gap of Cd$_x$Hg$_{1-x}$Te as a function of temperature and composition is given by the empirical equation [10]:

$$E_g = -0.302 + 1.93x - 0.81x^2 + 0.823x^3 + 5.32 \cdot 10^4 (1-2x)\left[(-1822 + T^3)/(255.2 + T^2)\right] \quad (1).$$

. The heavy holes effective mass is taken as $m_{Phh} = 0.43 m_0$, where $m_0$ is the free electron mass, and for the electron effective mass we assume the following expression which takes into account the band nonparabolicity (see, e.g., [11]):

$$m_e = m_0 \left(1 + 2 \frac{2m_0 P^2 (E_g + 2\Delta/3)}{\hbar^2 E_g (E_g + \Delta)}\right)^{-1} \quad (2),$$

where $P = 0.83 eV \cdot nm$ is the interband matrix element and $\Delta = 1.0 eV$ is the spin- orbit interaction constant.

Main dark current mechanisms in MCT photodiode structure are the diffusion current, band-to-band current (BTB), trap-assisted tunnelling (TAT) and Shokley-Reed-Hall (SRH) thermal processes. The latter two processes occur through the traps in the band gap. They both depend on the traps occupation factor. For that reason, taking into account only these two processes independently one can greatly overestimates the currents, and it is necessary to take into account the carriers balance at each trap. This approach, called a balance equation method, was developed in [12, 13] and was successfully applied in [9] for the description of experimental data for MCT and lead-tin-telluride photodiode structures [7, 9 ,14 ,15]. On this method we will dwell below.

To obtain the total recombination rate, two approximations were used: (*i*) the constant barrier field (valid for abrupt p-n-junctions as confirmed from capacitance-voltage characteristics for HgCdTe diodes), and (*ii*) constant quasi Fermi level across the barrier (is valid for forward and small reverse biases). Using those both approximations, one has the density of conduction band electrons and valence band holes for $0 \leq x \leq W$:

$$p(x) = N_A \exp[-q(V_0 - V)(W - x)/kTW], \quad n(x) = N_D \exp[-q(V_0 - V)x/kTW], \quad (3)$$

where $V_0$ is a built-in potential, $V$ is the reverse applied potential, $W = \sqrt{2\varepsilon_r\varepsilon_0(V_0-V)\frac{N_A+N_D}{qN_AN_D}}$ is the thickness of the n-p- junction, q is the electron charge, $N_a, N_d$ are acceptor and donor concentrations, $\varepsilon_0$ is the dielectric constant, and $\varepsilon_r = 17.5$ is the static dielectric constant.

Tunnelling rate characteristics were calculated in the *k-p*-approximation. The rate of carrier tunnelling from trap centers with energy $E_t$ above the valence band is given by

$$\omega_{c,v}N_{c,v} = \frac{\pi^2 qFm^*}{h^3(E_g-E_t)} \times |W_c|^2 \times \exp(-2\theta_{c,v}) \tag{4}$$

where $|W_c|^2$ is the square of the transition matrix element, and *F* is the electric field strength. In our calculations we have used the value $|W_c|^2 = 3\times10^{-67}$ J$^2$ cm$^3$, experimentally found [16] for Au in Si; actually, the magnitude variation of $|W_c|^2$ has only a minor effect on the current in comparison with the exponential term (see Equation 4). The parameter $\theta$ entering Eq. (4) is given by:

$$\theta = \int_{x_1(E)}^{x_2(E)} \mathrm{Im}(k_x(E))dx, \quad \text{where } k_x = \sqrt{\frac{3}{2}\frac{1}{P}}\sqrt{E(E_g-E)} \tag{5}$$

is the wave vector along the *x* direction, and $x_1$, $x_2$ are the endpoints of the tunnelling process. In the constant field approximation for the barrier potential, i.e., when $E_v(x)=q(V_o-V)x/W$ for $0\leq x \leq W$, the exponential factor is

$$\theta_{c,v} = \frac{\sqrt{3}WE_g^2}{8\sqrt{2}Pq(V_0-V)}\left[\frac{\pi}{2}+\arcsin(\pm 1 \mp 2\alpha)\pm 2(1-2\alpha)\sqrt{\alpha(1-\alpha)}\right], \tag{6}$$

where $\alpha = E_T/E_g$, and $\theta_c$ and $\theta_v$ are for conduction to deep level tunnelling and for deep band tunnelling processes respectively. For detailed balance considerations in *p-n*-junction we have used the following set of equations describing trap level capture (*c*) and emission (*e*) rates:

$$c_n^{tun} = n(x-x_t)[N_t-n_t(x)]\omega_c, \quad e_p^{tun} = n_T N_c \omega_c \quad \text{for tunnelling via conduction band;}$$

$$c_p^{tun} = p_{lh}(x+x_t)n_T(x)\omega_v, \quad e_p^{tun} = [N_T-n_T(x)]N_{vlh}\omega_v \quad \text{for tunnelling via valence band;}$$

$$c_n^{th} = n(x)[N_T-n_T(x)]\gamma_c, \quad e_n^{th} = n_T(x)n_1\gamma_c \quad \text{for thermal traffic via conduction band;}$$

$$c_p^{th} = pn_T(x)\gamma_v, \quad e_p^{tun} = [N_t-n_T(x)]p_1\gamma_v \quad \text{for thermal traffic via valence band} \tag{7}$$

Here $x_T=x_1-x_2$ is the tunnelling distance. Superscripts "tun" and "th" indicate tunnel and thermal processes respectively, $N_T$ is the density of traps, $n_T(x)$ is the density of electrons on traps at distance *x*, $p_{lh}(x)$ is the density of light holes, $N_c \cong N_{vlh} = 2(3E_g kT/8\pi P^2)^{3/2}\cdot(1+4kT/qE_g)$ are the conduction band ($N_c$) and light hole valence band ($N_{vlh}$) density of states respectively, $N_v \cong N_{vhh} = (2\pi m_{phh}kT)^{3/2}/(4\pi^3\hbar^3)$ is the valence band effective density of states deduced from a usual parabolic dispersion law. Temperature dependence of the thermal emission rates is given by parameters $n_1 = N_c \exp[-(E_g - E_T)/kT]$, and $p_1 = N_v \exp[(-E_T)/kT]$. Here $\omega_{c,v}$ are the tunnel capture constants from Eq. (4) and $\gamma_{c,v} = (N_T \tau_{e,p})^{-1}$ are the thermal capture constants for transitions from the conduction (valence) band respectively. Depending on the relations between $E_g$, $E_t(x)$ and $(V_0-V)$, the carrier at coordinate $x$ can or cannot tunnel to the point with the coordinate $(x+x_T)$ (in the valence band) or $(x-x_T)$ (in the conduction band), where $x_T = \frac{E_g - E_t}{q(V_0 - V)} W$ is the tunnelling distance in constant barrier potential approximation. Taking into account that

$$n(x - x_T) = n(x)\exp(\frac{E_g - E_T}{kT}), \quad p_{lh}(x + x_T) = p_{lh}(x)\exp(\frac{E_T}{kT}), \tag{8}$$

the non-equilibrium recombination rate for the case when there is a contribution to tunnelling via the conduction band $(E_g \leq E \leq q(V_0-V)+E_T)$ is given by:

$$U_a = \frac{N_T \gamma_v n_i^2 (\gamma_c + \omega_c \exp(\frac{E_g - E_t}{kT}))(\exp(-\frac{qV}{kT})-1)}{(p(x)+p_1)\gamma_v + (n(x)+n_1)\gamma_c + (n(x)\exp(\frac{E_g - E_T}{kT}) + N_c)\omega_c}. \tag{9a}$$

For the case when only the SRH generation/recombination transitions exist $(q(V_0-V) \leq E \leq E_g)$:

$$U_b = \frac{N_T \gamma_c \gamma_v n_i^2 (\exp(-\frac{qV}{kT})-1)}{(n(x)+n_1)\gamma_c + (p(x)+p_1)\gamma_v}. \tag{9b}$$

For the case when tunnelling transitions are possible for the valence band $(E_T \leq E \leq q(V_0-V))$:

$$U_c = \frac{N_T \gamma_c n_i^2 (\gamma_v + \omega_v \exp(\frac{E_t}{kT}))(\exp(\frac{qV}{kT})-1)}{(p_{lh}(x)\exp(\frac{E_T}{kT}) + N_{vlh})\omega_v + (p(x)+p_1)\gamma_v + (n(x)+n_1)\gamma_c}. \tag{9c}$$

And for the case when tunnelling is possible both from the valence band to the trap level and from the trap level to the conduction band $(E_g < E+E_T < q(V_0+V))$ one can obtain:

$$U_d = \frac{N_T n_i^2 (\omega_v \omega_c \exp(\frac{E_g}{kT}) + \omega_c \gamma_v \exp(\frac{E_g - E_t}{kT}))(\exp(-\frac{qV}{kT})-1) + \omega_v \gamma_c \exp(\frac{E_T}{kT}) + \gamma_c \gamma_v}{(p_{lh}(x)\exp(\frac{E_T}{kT}) + N_{lh1})\omega_v + (n(x)+n_1)\gamma_c + (p(x)+p_1)\gamma_v + (n(x)\exp(\frac{E_g - E_T}{kT}) + N_c)\omega_c}. \tag{9d}$$

The generation-recombination current is obtained by integrating the recombination rates across the p-n junction region:

$$I = q\left\{\int_0^{x_1} U_c dx + \int_{x_1}^{x_2} U_d dx + \int_{x_2}^{W} U_a dx\right\}, \quad \text{when} \quad q(V_0+V) > E_g$$

$$I = q\left\{\int_0^{x_1} U_c dx + \int_{x_1}^{x_2} U_b dx + \int_{x_2}^{W} U_a dx\right\}, \quad \text{when} \quad q(V_0+V) > E_g - E_T, E_T,$$

$$I = q\left\{\int_0^{x_{21}} U_c dx + \int_{x_2}^{W} U_b dx\right\} \quad \text{when} \quad E_T \leq q(V_0+V) < E_g - E_T,$$

$$I = q\left\{\int_0^{x_1} U_b dx + \int_{x_1}^{W} U_a dx\right\}, \qquad \text{when} \quad E_g - E_T \leq q(V_0 + V) < E_T, \qquad (10)$$

where $x_1 = \frac{E_g - E_t}{q(V_0 + V)} W$, $x_2 = \left(1 - \frac{E_t}{q(V_0 + V)}\right) W$ are the endpoints of the tunnelling range.

Other dark current mechanisms, such as band-to-band tunnelling, diffusion, SRH generation-recombination currents in quasi neutral $n^+$- and $p$-regions of the diode, radiative recombination in the depletion region and quasi neutral $n^+$- and $p$-regions, and the Auger recombination are considered as additive and independent terms. The largest contributions to the dark current of the diode are made by the diffusion component, by the SRH component for the holes and Auger 1 recombination processes in the quasi-neutral region of the $p-n-$ junction, and by BTB tunnelling.

For band-to-band tunnelling we have used the constant field approach (as well as for the TAT before) in the Wentzel- Kramers –Brillouin (WKB) approximation for the two-band dispersion rule with allowance made for nonparabolicity of the band spectrum [17]:

$$I_{BTB} = -\frac{(q\varphi - E_g/2)q^3 F}{\sqrt{2}\pi^3 \hbar^2}\left(\frac{m_y^* m_z^*}{E_g m_x^*}\right)^{1/2}\left\{E_4\left[\left(\frac{q\varphi}{q\varphi - E_g/2}\right)^{1/2} K\right] - \left(\frac{E_g}{2q\varphi} - E_g\right)^{3/2} E_4\left[(2q\varphi/E_g)^{1/2} K\right]\right\}, \quad (11)$$

where $x$ is the direction of the current, and $m_i^*$, $i=x,y,z$ are the principal components of the effective mass tensor (for MCT all the three components are equal to $m_e^*$), $K = \pi/2qF\hbar(m_x E_g^3)^{1/2}$, $\varphi = V_0 + V$ is the junction potential, $F = (V_0 + V)/W$ is the electrical field strength in the junction, and $E_4(c) \equiv \int_1^{\infty} e^{-Cx} x^{-4} dx$ is an exponential integral.

Auger 1 generation /recombination processes is significant only in $n$ quasi neutral region of the photodiode structure. This current is described as:

$$J_{A1\_n} = 1/2 q \tau_{A1}^{-1} N_D L_p (\exp(qV/kT) - 1), \qquad (12),$$

where $\tau_{A1}$ is the intrinsic Auger1 recombination time:

$$\tau_{A1} = \frac{3.8 \cdot 10^{-18} \varepsilon_\infty^2 (1+\beta)^{1/2}(1+2\beta)\exp\left[\left(\frac{1+2\beta}{1+\beta}\right)\frac{E_g}{kT}\right]}{(m_e/m_0)|F_1 F_2|^2 (kT/E_g)^{3/2}},$$

$\beta = m_e/m_{phh}$, $\varepsilon_\infty$ is the high frequency dielectric constant and $|F_1 F_2|^2 \approx 0.1^2 \div 0.3^2$ is a square of the electron wave functions overlap integral.

The SRH generation – recombination currents in quasi-neutral $n$ and $p$ regions are essential at small biases and are comparable with the diffusion current (in the approximation of infinitely wide $n$ or $p$ regions):

$$I_{SRHn,p} = qN_{tv}\gamma n_i^2 L_{n,p}/N_{d,a} *(\exp(qV/kT) - 1) \qquad (13),$$

where $\gamma = (N_{tv}\tau_v)^{-1}$ is the capture constant, and $L_{n,p} = \sqrt{kT\mu_{n,p}\tau/q}$.

Results of fitting procedure for dark currents and incremental resistance are shown at the Figure 3. In both cases for dark currents and differential resistances there is a good agreement with experimental results. It is seen from Figure 3 that at weak reverse bias the major current for investigated structures is the diffusion current (including non-tunnelling components such as the current of SRH recombination outside the depletion region of the $n^+$-$p$- junction, Auger and radiative recombination currents). From this region of the current-voltage curve the concentrations of donor and acceptor impurities $N_d$ and $N_a$, the lifetimes of the carriers outside the depletion region $\tau_{pv}$, $\tau_{nv}$, and the trap concentration in the band gap outside the depletion region $N_{tv}$ were found. The last one should be of the same order as the acceptor concentration, since the traps are created during the growth and annealing processes.

At higher absolute value of the reverse bias voltage the processes of tunnelling are strengthened. Both TAT and BTB tunnelling strongly depend on the geometry of the $p$-$n$-junction, and thus on the donor and acceptor concentrations. Therefore, the concentrations which result from the fitting procedure should match the diffusion current at low biases, and at high biases they should match tunnel currents. In the Figure 4 it is shown how the fitting parameters change with the increase of γ-radiation exposure.

It was assumed that the traps in the band gap are related to the donor-type interstitial Hg with the energy $E_t$ = 0.7 $E_g$ above the top of the valence band and the concentration $N_t$ = 1.5·10$^{16}$ cm$^{-3}$ that is lower than the donor concentration $N_d$ = 1·10$^{17}$ cm$^{-3}$ in $n^+$ region and higher than the acceptor concentration $N_a$ = 1.35·10$^{15}$ cm$^{-3}$ in $p$ region.

Under boron ion implantation, the $n^+$-$p$-junction in MCT layers is formed by the diffusion of Hg$^+$ interstitials from the implant damage region near the surface into the MCT layer and annihilation with Hg vacancies [18,19]. As a result, the doping profile of the diode structure is $n^{++}$-$n^+$-$p$, where the $n^{++}$ region has aproximately 1 μm depth near the surface, which is roughly the same as the depth of the implant damaged region in the wide-gap layer of $Hg_{0.35}Cd_{0.65}Te$ in the structures studied. Actually, the $n^+$-$p$-junction is formed at the depth of about $3 \div 4$ μm, where the $n^+$-region Hg vacancies have been annihilated with interstitial Hg donors and the $p$-region remains doped by acceptor-like $Hg^{2-}$ vacancies.

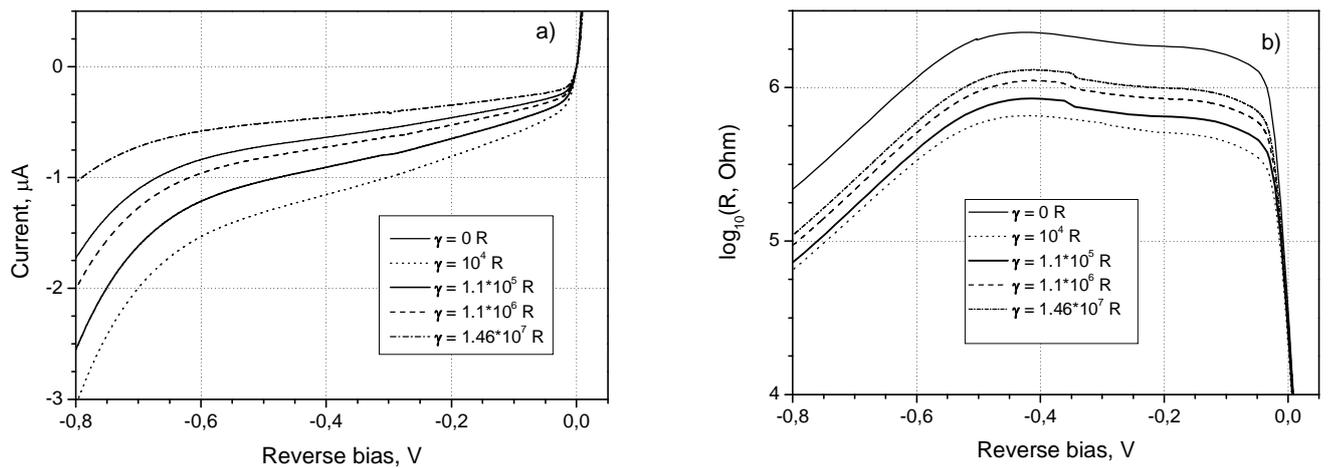

Figure 3. Calculated dark current – voltage and differential resistance – voltage characteristics for Hg$_{1-x}$Cd$_x$Te (x=0.2236) photodiodes. The styled lines correspond to the results of modeling for the following values of free parameters: $E_t$ = 0.7 $E_g$, $N_d$ = 1·10$^{17}$ cm$^{-3}$, $N_a$ = 1.35·10$^{15}$ cm$^{-3}$, $N_t$ = 1.5·10$^{16}$ cm$^{-3}$, $N_{tv}$ = 1.2·10$^{16}$ cm$^{-3}$, τ = 3.7·10$^{-9}$ s, τ$_v$ = 3·10$^{-9}$ s before γ radiation exposure. The change of parameters $N_t$, $N_{tv}$, τ, and τ$_v$ with the radiation dose is shown in Figure 4.

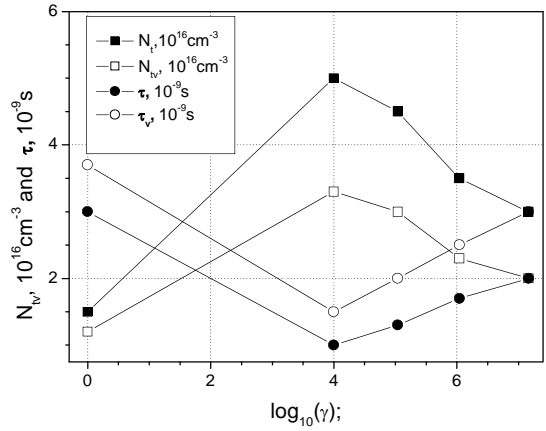

Figure 4. The dependence of the trap lifetime
and concentration on γ-radiation exposure.

Under γ-ray treatment with the lowest dose $10^4$ R the trap concentration first increases with the carrier lifetime decrease. At higher doses the concentration of traps decreases (see Figure 4), which can be related to the decrease of the concentration of interstitial Hg.

For the second group of diodes, this assumption agrees well with experiments on X-ray induced photoemission [20], where the existence of two types Hg atoms in MCT layers was revealed: one in the bulk, and the other forming metallic clusters at the surface layer. For almost all of the metallic clusters the binding energies of core level electrons are higher than for the corresponding bulk material.

In our case, for the second group of diodes, fabricated by the same technological process, it seems that γ-radiation "sweeps" the already existing interstitial Hg to the surface and simultaneously creates new interstitial and vacancy defects. Thus, at low doses, the concentration of the electrically active interstitial Hg defects increases. At higher doses, as the initial "wave" of the interstitials reaches the surface and forms clusters, the concentration of interstitial Hg in the bulk decreases, stabilizing at some lower value since the process of creating new defects is suppressed by the increased probability of annihilation with vacancies. At the moment, it is not clear why there exist two groups of diodes obtained by the same technology from the same crystal.

We investigated the radiation hardness of readout integrated circuits (ROIC) for MCT photodiodes as well. The operability was examined by measuring the following parameters: output signals amplitude, output voltage dynamical range, bias voltage spread, charge transfer inefficiency, charge voltage transmissive characteristic linearity, charge capacity, and power consumption. It was revealed that the ROIC fail after irradiating with smaller doses as compared with photodiodes above described. The critical doses were about $10^5$ R. The main parameter that varies with the increase of exposure is the direct injection transistor threshold voltage, but up to $10^5$ R exposures the ROIC still remained fully functional at operating temperature $T$=80 K.

## 4.     Summary

The problem of radiation hardness is important for the determination of operational characteristics of IR-sensors for many applications. Exposing photodiode arrays to various doses of γ-radiation ($1·10^4$ R, $1.1·10^5$ R, $1.1·10^6$ R and $1.46·10^7$ R), we have found that one can distinguish two groups of photodiodes by their response to the radiation treatment. It should be emphasized that both groups of diodes belonged to the same array, so they were fabricated on the same crystal under identical technological conditions. Most of the diodes (nearly 70%) do not change their operational characteristics such as current-voltage characteristics and differential resistance even when exposed to γ-radiation with the highest dose. The remainder of the diodes showed slightly decreased incremental resistances and larger currents under reverse biases after γ-radiation exposure. Those parameters do not change linearly with the radiation dose: instead, they demonstrate strong worsening at small exposure and then a recovery with the increase of the dose. But even for the "worst" photodiodes, their operational characteristics, such as dark currents and incremental resistance after γ radiation exposure, are quite acceptable for the use in FPAs with proper readouts.

Within the framework of the balance equation model, the current mechanisms for MCT photodiode arrays exposed to various doses of γ-radiation can be satisfactorily explained by the change of trap concentrations and carrier lifetimes in the depletion region and in the quasi-neutral region. We argue that the decrease of current at larger doses of γ radiation can be related to the decrease of concentration of Hg interstitial tunnelling centers in the gap of MCT due to their clustering.

## 5. Acknowledgements